\documentclass[12pt]{article}

\usepackage{graphicx}
\usepackage{dcolumn}
\usepackage{bm}
\usepackage{subfigure}

\begin{document}

\title{TRANSFORMATION KINETICS ON MECHANICAL ALLOYING}
\author{Igor F. Vasconcelos\footnote{Corresponding author. E-mail: idevasco@nd.edu} \\ 
\textit{Department of Physics, University of Notre Dame,} \\
\textit{225 Nieuwland Science Hall, 46556, Notre Dame, IN, USA} 
\and Reginaldo S. de Figueiredo \\ 
\textit{LG. Philips Displays - GDE - TZ 1, Zwaanstraat 2,} \\
\textit{PO Box 807, 5600 AV Eindhoven, The Netherlands} }

\maketitle

\begin{abstract}
We propose and develop a model as an attempt to describe the local mechanism of mechanical alloying. This model is based on the 
observation of global parameters, such as the volumes of the various phases present in the material, and their transformations 
during the process. The model is applied to milled Fe-Cu and Fe-N experimental results obtained in previous 
works~\cite{vasconcelos99,foct94}. In the milling of Fe-Cu system, four stages, whose features are in accordance with experimental 
observation, are identified:(i) energy storage; (ii) initiation of reaction and rate increase; (iii) decrease of the reaction 
rate and (iv) stationary stage. No energy storage stage is observed and no saturation is achieved in the milling of Fe-N system. 
Moreover, we propose a modification on the global kinetic law introduced in~\cite{matteazzi93} based on the different
nature of Fe-Cu and Fe-N systems. This modification allows the extension of the validity of this law. \\
\end{abstract}

\section{Introduction}

In classical studies of phase transformations, kinetic laws allow a global description of volume phase as a function of time 
evolution thus distinguishing them from growth laws. However, growth laws translate a local microscopic modeling of phase 
transformation phenomena and provide the construction of kinetic laws from microscopic observation and (or) models. Both approaches 
exist in mechanical alloying (MA) although they are rarely explicit in the literature. 

MA is a cheap, efficient, and flexible technique (see the survey by Suryanarayana~\cite{surya01}) which has attracted much attention from 
scientists of various research areas. Recently, MA has been used to produce materials for biological application~\cite{silva02} and to 
investigate the spin-glass character of nanograin boundaries~\cite{navarro99}, amongst many other applications. Although much of the 
research is based on empirical conclusions obtained directly from experimental results, there has been some attempts to make a quantitative 
description of MA. A quantitative description is very important and necessary in order to provide a way to establish predictable capabilities 
for the MA processes~\cite{cocco00}. 

Abdellaoui and Gaffet~\cite{abdellaoui94,abdellaoui95} have shown that neither the shock energy nor the shock frequency govern the 
Ni$_{10}$Zr$_{7}$ amorphization process in a planetary ball mill if taken into account separately. Only the injected shock power is 
responsible for the ball milled final product. In another study of Ni-Zr amorphization, Chen~\textit{et al.}~\cite{chen93}, using a 
vertical vibrating frame grinder, have found that these compounds become fully amorphized if a defined milling intensity ($I$) is 
greater than a threshold value. The milling intensity is defined as:
\begin{equation}
I=\frac{M_bV_{max}f}{M_p} \textrm{,} \label{eqintensity}
\end{equation}
where $M_b$ is the ball mass, $V_{max}$ is the maximum velocity of the ball, $f$ the impact frequency, and $M_p$ the mass of the
powder in the vial. Therefore, according to~\cite{chen93}, the same final product can be achieved by transfering large momentum
with low frequencies or small momentum with high frequencies. These two approaches, while successful in describing individual processes in
particular types of mill, have some built in difficulties that prevent them to fully describe a general MA experiment. The shock power
proposed in~\cite{abdellaoui94,abdellaoui95} is a function of balls' kinetic energy and impact frequency. However, in a planetary mill,
both parameters are functions of the velocity of the balls in the sense that the faster a ball moves, the higher its kinetic energy and 
the more times it will suffer impacts. Moreover, in the vertical vibrating mill used to investigate the role of the milling 
intensity~\cite{chen93}, the shock frequency is determined by the inverse of the time taken by ball to move up and down. This time
is proportional (and as a consequence the frequency is inversely proportional) to potential energy. The velocity $V_{max}$ is itself
proportional to potential energy which makes the product $V_{max}f$ close to a constant value. These two approaches also neglect
the importance of the material's response to the impacts.

MA is a ``far from equilibrium technique'' and despite its dynamic characteristics Yavari \textit{et al.}~\cite{yavari92,yavari94} 
have employed equilibrium thermodynamics to explain the MA mechanism. Pure elements form nanocrystalline structures after severe 
mechanical deformation and the energy stored in the large grain boundaries could be the driving force during alloying. Examining 
the immiscible Fe-Cu system prepared by mechanical alloying, Yavari \textit{et al.} have shown that the available interfacial energy is 
not enough to compensate the positive enthalpy of mixing. However, repeated fractures break the particles into very small fragments 
with high free energies which dissolve themselves into the matrix by the Gibbs-Thomson effect. 

Matteazzi \textit{et al.}~\cite{matteazzi93} have proposed a global kinetic law for MA of iron carbides: 
\begin{equation} 
V(t)=ke^{-bt^c} \textrm{.} \label{eqmatteazzi} 
\end{equation} 
where $V(t)$ is the transformed phase volume and $k$, $b$ and $c$ are empirical constants.  They have proposed that $c$ is connected 
to impact energies and $b$ is related to ``work conditions''. The value of $k$ sets the scale. 

In a previous work~\cite{vasconcelos99}, we reported a discussion about the MA process based on experimental and numerical results. 
A nanocrystaline fcc-FeCu solid solution was obtained by MA and analyzed by x-ray diffraction, M\"{o}ssbauer spectroscopy, and 
differential calorimetry. In addition, simulations of the material behavior during uniaxial deformations, compatible with what
we expect on an impact were performed using molecular dynamics. 
We claimed that the number of balls in the mill vial, which is related to impact frequency, did not affect the energy transfer 
processes, but rather only the reaction rates. This finding has recently been confirmed by Girot \textit{et al.}~\cite{girot00}. 
We also showed that a local description of MA process is necessary in order to understand its driving mechanisms and that many factors 
such as; powder deformation by the milling tools, chemical and physical properties of materials, strain/stress fields, and collision 
energies have to be taken in account together to describe MA. This view was recently raised in the review by Takacs~\cite{takacs02}.

In this contribution we propose and develop a model as an attempt to describe the local mechanism of MA. This model is based on the 
phase volume transformations during the process and is applied to milled Fe-Cu and Fe-N experimental results obtained 
previously~\cite{vasconcelos99,foct94}. 

\section{The Model} 

Our first step to understand MA mechanisms is to investigate how the phase transformations are driven in the course of the process. 
This can be done using measured global parameters such as the phase volumes. If a given phase $\alpha$ occupies a volume $V_{\alpha}(t)$ 
in the total volume $V$, then the concentration $X_{\alpha}(t)$ is given by: 
\begin{equation} 
X_{\alpha }(t)=\frac{V_{\alpha }(t)}{V}\textrm{.} 
\end{equation} 
 
MA is essentially a discrete process characterized by discrete energy transfers taking place at individual impacts~\cite{delogu00}, which
means that only a small fraction of the powder is processed at each event. Moreover, a particle of powder under an impact has no memory from the 
previous impacts~\cite{vasconcelos99}. Significant powder transformations occurring during milling impacts are confined in a small volume $v$ which 
is the powder volume trapped by the milling tools, as suggested in~\cite{girot00}. The $\alpha$ phase volume in $v$ is $v_{\alpha}$ and assuming a 
homogeneous powder~\cite{lecaer02}, the relative concentration is then given by: 
\begin{equation} 
X_{\alpha }(t)=\frac{v_{\alpha }(t)}{v}\textrm{.} 
\end{equation} 

Let us consider that $\alpha$ phase is created after a MA driven solid state reaction between two other phases $\beta$ and $\gamma$. 
The volume created or transformed $\Delta v_{\alpha}$ after an impact involving a $\beta \gamma $ interface is a function of the 
probability to find a $\beta \gamma $ phases reaction ($P_{\beta \gamma }(t)$) and tribochemical activity 
($\xi_{\beta \gamma }^{\alpha }(t)$) given by: 
\begin{equation} 
\Delta v_{\alpha }(t)\longrightarrow F\left( P_{\beta \gamma }(t),\xi 
_{\beta \gamma }^{\alpha }(t)\right) \textrm{.}  \label{eqVolImpacto} 
\end{equation} 
The word tribochemical includes many aspects such as chemical activities, morphology of components and impact characteristics 
(energy, frequency, etc.). In a first approximation eq.~\ref{eqVolImpacto} can be written as: 
\begin{equation} 
\Delta v_{\alpha }(t)=\sum_{\beta \gamma }P_{\beta \gamma }(t)\xi _{\beta 
\gamma }^{\alpha }(t)v\textrm{.} 
\end{equation} 
This summation is carried out over all $\beta \gamma $ phase reactions which could possibly lead to production of $\alpha$.  

A formal description of $\xi _{\beta \gamma}^{\alpha }(t)$ is quite difficult; however, it is possible to describe the volume of a 
phase during the total milling time $\Delta t$. The evolution of the $\alpha$ phase global volume ($\Delta V_{\alpha}(t)$) in the 
course of time interval $\Delta t$ is given by: 
\begin{equation} 
\Delta V_{\alpha }(t)=\Delta v_{\alpha }(t)\times f\Delta t =  
\sum_{\beta \gamma }P_{\beta \gamma }(t)\xi _{\beta 
\gamma }^{\alpha }(t)vf\Delta t\textrm{,}  \label{eqEvolVolTotal} 
\end{equation} 
where $f$ is the impact frequency and $f\Delta t$ is the number of impacts during the time interval $\Delta t$. 

The probability to find a $\beta \gamma $ interface which can react is given by the relative surface areas of the $\beta $ and $\gamma$  
($S_{\beta }(t)$ and $S_{\gamma }(t)$) phase grains  in the volume $v$: 
\begin{equation} 
P_{\beta \gamma }(t)=\frac{S_{\beta }(t)S_{\gamma }(t)}{S_{T}^{2}}\textrm{,} 
\label{eqProbAreas} 
\end{equation} 
where $S_{T}$ is the total surface area in the volume $v$. Considering spherical grains, eq.~\ref{eqProbAreas} can be written as: 
\begin{equation} 
P_{\beta \gamma }(t)=\frac{X_{\beta }(t)X_{\gamma }(t)}{r_{\beta }r_{\gamma }%
}\left( \sum_{i}\frac{X_{i}(t)}{r_{i}}\right) ^{-2}  \label{eqProbInterm} 
\end{equation} 
where $r_{i}$ is the particle radius of phase $i$ and the summation is carried out over all the phases in the volume $v$. 

This analysis can be greatly simplified if we consider two-phased systems, composed by phases $\varepsilon $ and $\delta $. The possible 
reactions between the phases are 
\begin{equation} 
\varepsilon +\varepsilon \longrightarrow \Delta v_{\varepsilon }=0,\Delta v_{\delta }=0 \textrm{,}
\end{equation}
\begin{equation}
\delta +\delta \longrightarrow \Delta v_{\varepsilon }=0,\Delta v_{\delta}=0 \textrm{,}
\end{equation}
\begin{equation}
\varepsilon +\delta \longrightarrow \Delta v_{\varepsilon }<0,\Delta v_{\delta }>0\textrm{ or }\Delta v_{\varepsilon }>0,\Delta v_{\delta }<0\textrm{.}
\end{equation} 
The summation in eq.~\ref{eqEvolVolTotal} is reduced to just one term to translate the $\varepsilon $ (or $\delta $) volume phase evolution in 
such a way that: 
\begin{equation} 
\Delta V_{\varepsilon }(t)=P_{\varepsilon \delta }(t)\xi _{\varepsilon 
\delta }^{\varepsilon }(t)vf\Delta t\textrm{.} 
\end{equation} 
In this way, the tribochemical activity $\xi _{\varepsilon \delta}^{\varepsilon }(t)$ can be written as: 
\begin{equation} 
\xi _{\varepsilon \delta }^{\varepsilon }(t)=\frac{\Delta V_{\varepsilon }(t)}
{\Delta t}\frac{1}{P_{\varepsilon \delta }(t)vf}\textrm{.} 
\label{eqGeralTribo} 
\end{equation} 
Equation \ref{eqGeralTribo} presents at the same time a kinetic factor ($\frac{\Delta V_{\varepsilon }(t)}{\Delta t}$), a term of 
chemical nature ($P_{\varepsilon \delta }(t)$) and a ''condition of milling'' term ($vf$). The probability of a $\varepsilon \delta$ 
interface is given by eq.~\ref{eqProbInterm} and can be written as: 
\begin{equation} 
P_{\varepsilon \delta }(t)=\frac{X_{\varepsilon }(t)X_{\delta}(t)r_{\varepsilon }r_{\delta }}{\left[ X_{\varepsilon }(t)r_{\delta 
}+X_{\delta }(t)r_{\varepsilon }\right] ^{2}}\textrm{,}  \label{eqProbInterm1} 
\end{equation} 
and if the particles radii are comparable ($r_{\varepsilon }\approx r_{\delta }$), eq. \ref{eqProbInterm1} can be reduced to: 
\begin{equation} 
P_{\varepsilon \delta }(t)=X_{\varepsilon }(t)[1-X_{\varepsilon }(t)]\textrm{,} 
\label{eqProb} 
\end{equation} 
since $X_{\varepsilon}+X_{\delta}=1$. Furthermore, if the impact frequency is high, we get: 
\begin{equation} 
\frac{\Delta V_{\varepsilon }(t)}{\Delta t}\approx \frac{dV_{\varepsilon }(t)}{dt}=\frac{dX_{\varepsilon }(t)}{dt}V\textrm{.}  
\label{eqVarVol} 
\end{equation} 

Using eq.~\ref{eqProb} and \ref{eqVarVol} with eq.~\ref{eqGeralTribo} we obtain the proportionality relation: 
\begin{equation} 
\xi^{\varepsilon }(t)\propto 
 \frac{dX_{\varepsilon}(t)}{dt}\frac{1}{X_{\varepsilon }(t)[1-X_{\varepsilon }(t)]} \textrm{.} 
\label{eqTribo} 
\end{equation} 
At this point we drop the lower indices $\varepsilon$ and $\delta$ because they are no longer necessary. Equation~\ref{eqTribo}
describes the time evolution tribochemical activity to destroy (or create) the $\varepsilon$ phase. The same result applies to $\delta$.

\section{Fe-Cu System} 

We used the Fe$_{55}$Cu$_{45}$ system studied in a previous work \cite{vasconcelos99} to perform an analysis of phase volume 
evolution as described in eq.~\ref{eqEvolVolTotal}. We used a \textit{Fritsch Pulverisete 5} planetary ball mill to perform two 
sets of experiments, with 4 and 12 balls in the mill vial. The rotational speed used was 350 rpm for both experiments. 
The Fe-Cu samples were prepared by ball milling using commercial Fe and Cu 
metallic powders. In the course of milling, the iron atoms are incorporated into the fcc copper matrix so that at any time during
the process, there are fractions of the initial bcc Fe and fcc Cu and an amount of transformed fcc FeCu solid solution. The
observed trend is that of total incorporation of iron by the fcc matrix resulting on fcc FeCu alone~\cite{vasconcelos99,jiang93}.
Therefore, in the crystallographic point of view, the system consists of a bcc phase (Fe) that decreases gradually and a fcc
phase (Cu + FeCu). We applied the model proposed in this contribution to analyze the bcc phase's kinetics of transformation. 

\begin{figure} 
\begin{center} 
\includegraphics[angle=-90,width=\linewidth]{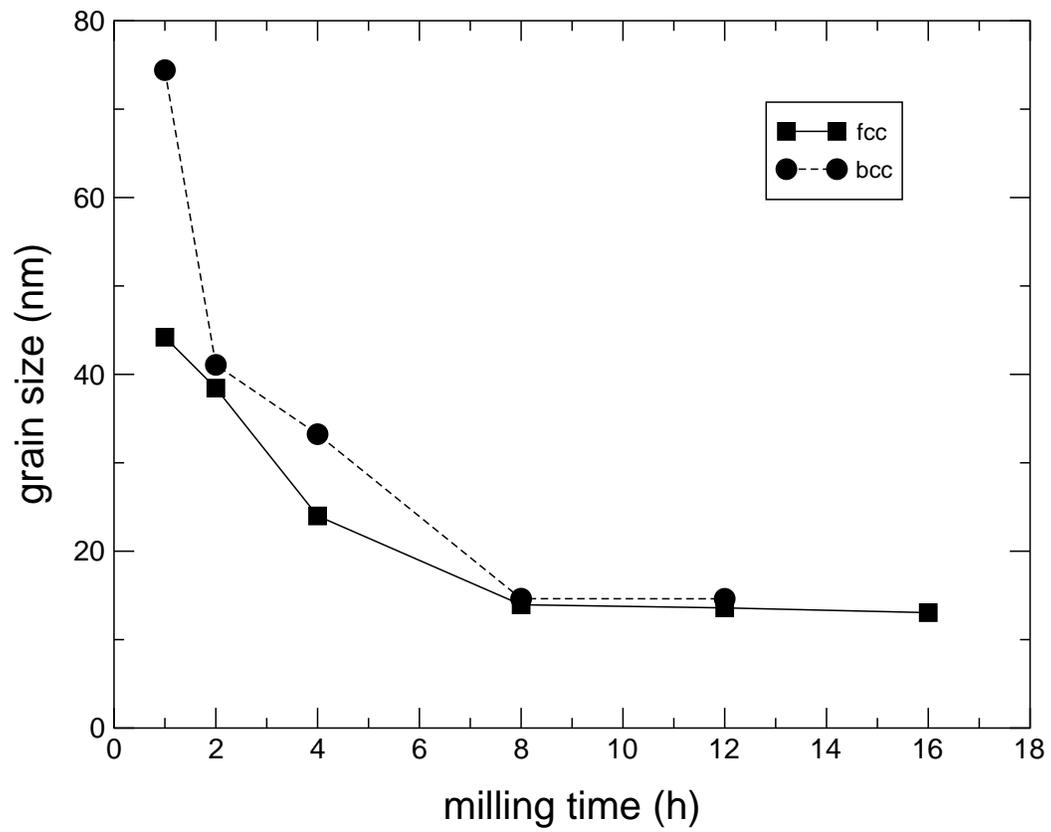}
\end{center} 
\caption{Grain size evolution of bcc and fcc phases during milling 
for Fe-Cu milled with 350 rpm and 4 balls revealing a similar 
behavior for the two phases.}\label{figTamGrao} 
\end{figure}

The grain size evolution presents a similar behavior for both bcc and fcc phases and $r_{bcc}\approx r_{fcc}$, (table~\ref{tabGrao} 
and fig.~\ref{figTamGrao}). Thus, the evolution of tribochemical activity to destroy the bcc phase can be followed by 
(eq.~\ref{eqTribo}): 
\begin{equation} 
\xi ^{bcc}(t)\propto \frac{dX_{bcc}(t)}{dt}\frac{1}{ X_{bcc}(t)[1-X_{bcc}(t)]} \textrm{.}  \label{eqTriboFeCu} 
\end{equation} 

\begin{figure} 
\centering
\subfigure[350 rpm -  4 balls] {
	\includegraphics[angle=-90,width=8cm]{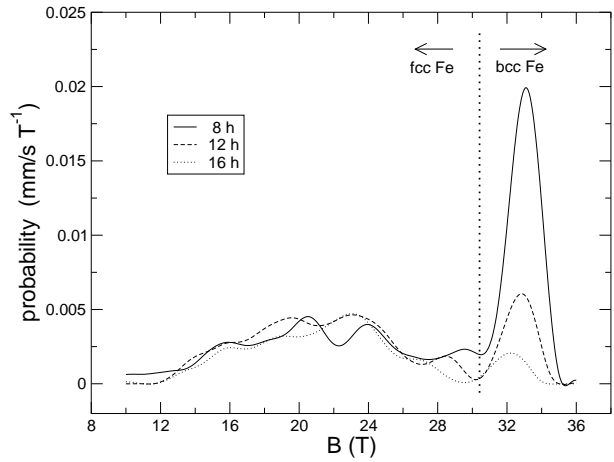} } \\
\vspace{0.5cm} 
\subfigure[350 rpm - 12 balls] { 
        \includegraphics[angle=-90,width=8cm]{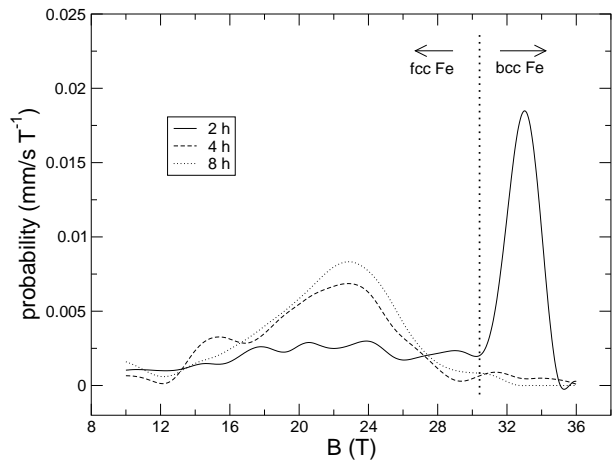} } 	
\caption{BHF distributions for some Fe-Cu milled samples showing a 
clear separation between Fe-fcc and Fe-bcc phases (low and high 
BHF values respectively)} 
\label{figDistrib} 
\end{figure}

\begin{figure} 
\begin{center} 
\includegraphics[angle=-90,width=\linewidth]{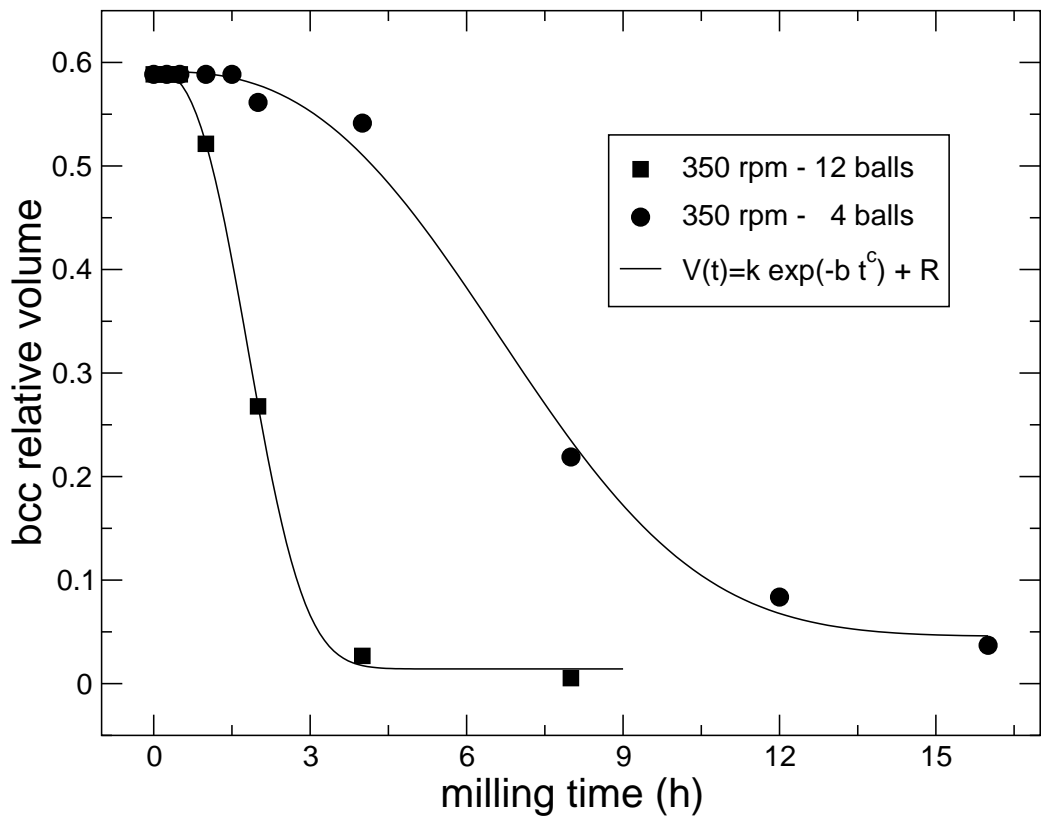}
\end{center} 
\caption{Bcc relative volume phase for the Fe-Cu system obtained 
by M\"{o}ssbauer spectroscopy.}\label{figVolume} 
\end{figure}

The evolution of phase concentration $X_{bcc}(t)$ can be deduced by the magnetic hyperfine field distributions (BHF) obtained by 
M\"{o}ssbauer spectroscopy (fig.~\ref{figDistrib}). Bcc concentration is the area under the spectra with BHF greater than 30 T. 
Figure~\ref{figVolume} shows the evolution curves for the bcc phase volume in the course of milling.

\begin{figure} 
\begin{center} 
\includegraphics[angle=-90,width=\linewidth]{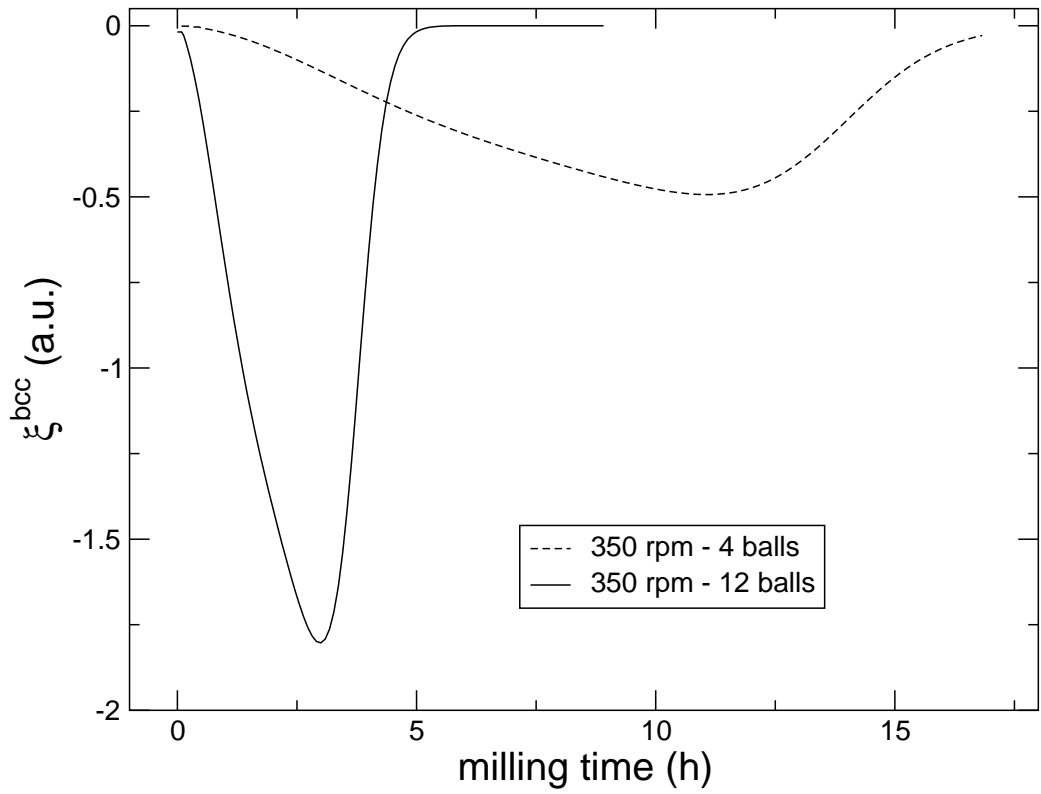}
\end{center} 
\caption{Tribochemical activity to destroy the bcc phase in the 
Fe-Cu system as a function of milling time. The curves were calculated
from data in Fig.~\ref{figVolume} using eq.~\ref{eqTriboFeCu}.}\label{figTriboses} 
\end{figure}

The tribochemical activities versus milling time for the two experiments described are shown in fig.~\ref{figTriboses}.  The curves were
calculated using eq.~\ref{eqTriboFeCu} and the data in Fig.~\ref{figVolume}. The 
scales, however, are not the same because impact frequencies, which are related to the number of balls, are different for each experiment.  
Figure~\ref{figTribo} shows $\xi $ vs. time for the two experiments rescaled by number of balls, in such a way that we get 
tribochemical activities per ball versus milling time per ball. 

\section{Fe-N System} 

\begin{figure} 
\begin{center} 
\includegraphics[width=\linewidth]{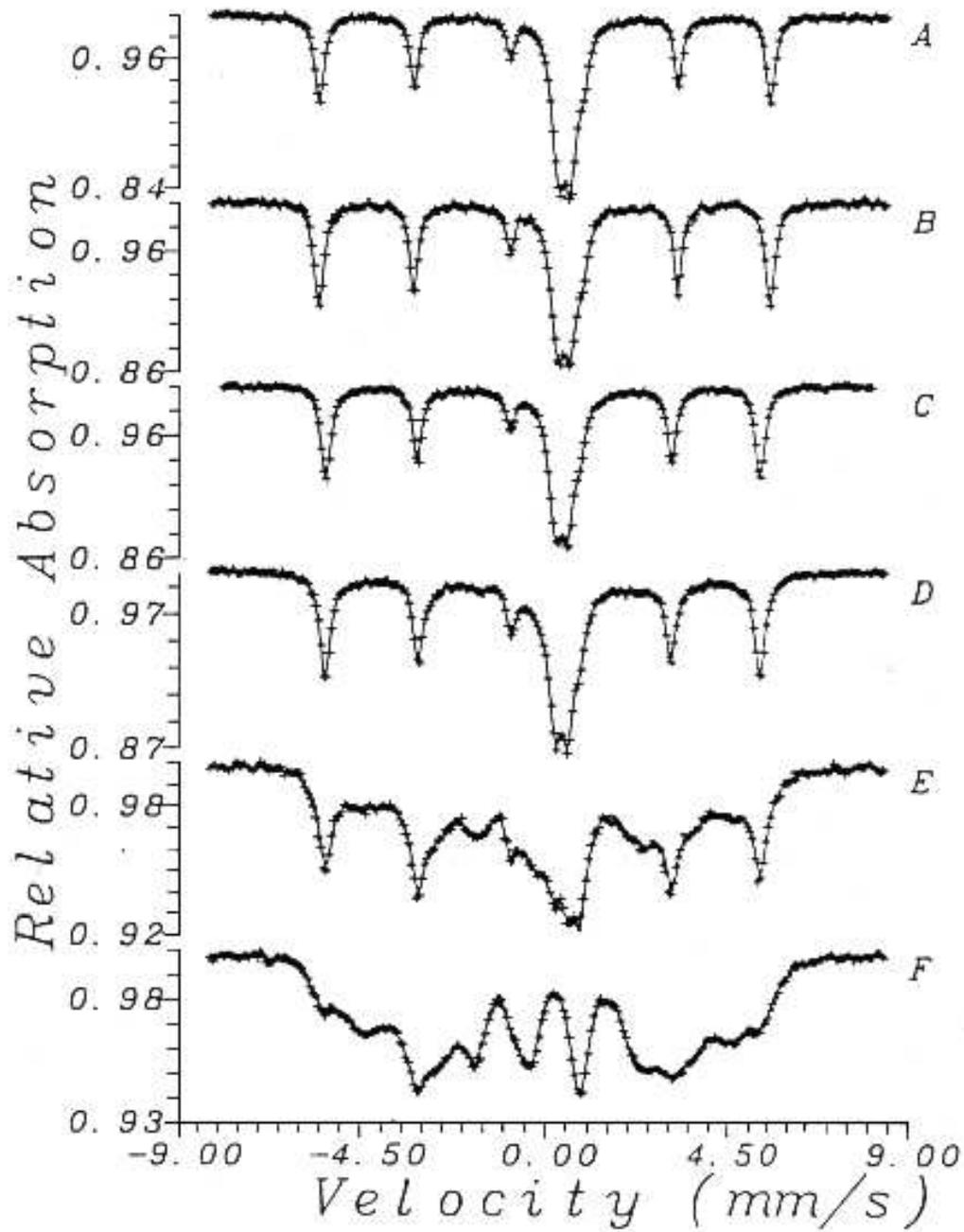}
\end{center} 
\caption{M\"ossbauer spectra at room temperature for milled $Fe_{2}N+2Fe$ up to (a) 5min; 
(b) 20min; (c) 1h; (d) 2h; (e) 4h and (f) 16h} \label{figMossNit} 
\end{figure}

\begin{figure} 
\begin{center} 
\includegraphics[angle=-90,width=\linewidth]{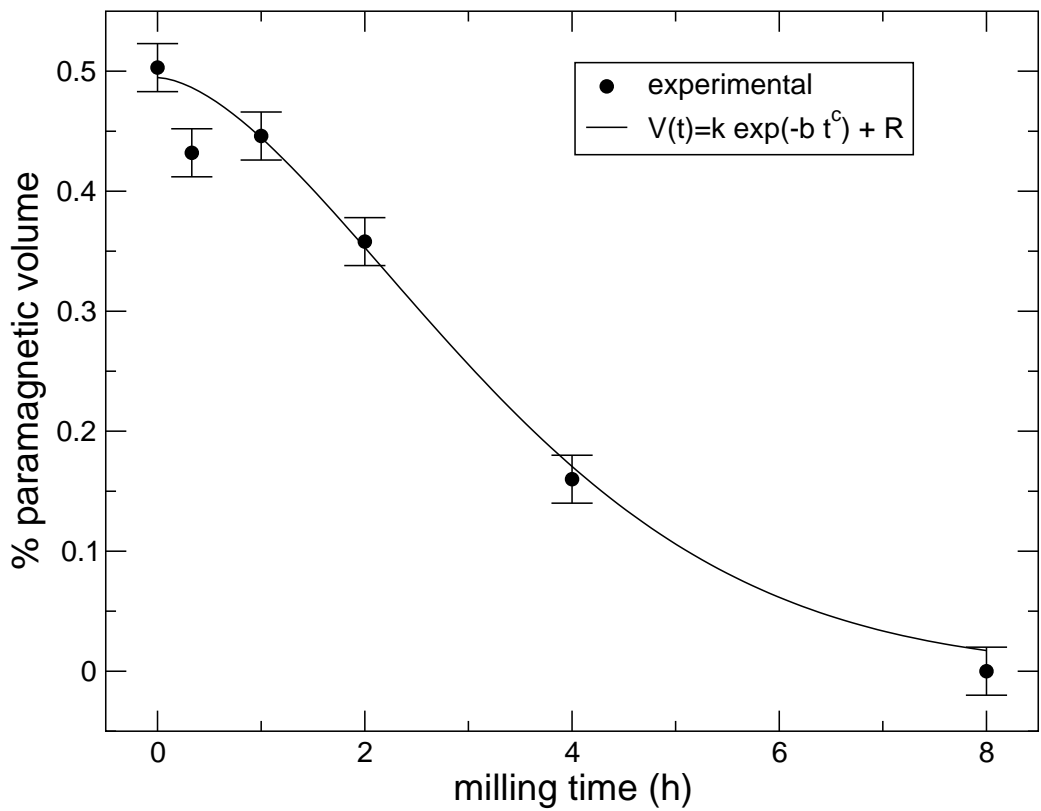}
\end{center} 
\caption{Paramagnetic relative volume phase for the Fe-N system obtained by 
M\"ossbauer spectroscopy} \label{figVolNit} 
\end{figure}

The $2Fe + Fe_{2}N \longrightarrow Fe_{4}N$ reaction obtained by MA studied in a previous work~\cite{foct94} is also a good 
application for the model discussed in this contribution. The transformation was followed through the volume fraction of $Fe_{2}N$ since 
the existence domain of this paramagnetic phase is relatively narrow. In addition to this, the system $2Fe+Fe_{2}N$ is two-phased, 
composed by a paramagnetic and a ferromagnetic phase; also the reaction product ($Fe_{4}N$) is ferromagnetic. Therefore, the 
phase volume of the paramagnetic $Fe_{2}N$ can be easily separated by M\"ossbauer spectroscopy. It is the area of the doublet subspectra 
of Fig.~\ref{figMossNit}. The $Fe_{2}N$ paramagnetic phase volume evolution is shown in Fig.~\ref{figVolNit}.

The Fe-N equilibrium phase diagram is composed by 4 crystallographic phases named $\alpha-Fe$, $\gamma'-Fe_4N$, 
$\varepsilon-Fe_{2.1-3.6}N$ and $\delta-Fe2N$. At high temperature one also finds a gamma phase $\gamma-Fe_XN$ and a epsilon nitrogen 
poor $\varepsilon-Fe_4N$. One can also find a metastable long range ordered phase $\alpha''-Fe_{16}N_2$. However, Fe-N 
phases obtained by milling are quite distinct. Milling of $\gamma'-Fe_4N$ and 2$Fe+Fe_2N$ leads to a hexagonal phase which 
presents different nitrogen distribution. Moreover, milling of $4Fe+Fe_4N$ leads to a hyperferrite instead of the well 
ordered $\alpha''-Fe_{16}N_2$~\cite{foct94,foct93,defigueiredo94,defigueiredo95,foct96,foct96a}. The large domain of epsilon phases, 
the wide range of the Nitrogen distribution, and the nanocrystalline character of these phases unable any attempt to distinguish  
nitrogen compositions in epsilon phases. What can be observed is that during most of the milling of $2Fe+Fe_2N$, the system is 
mainly composed of $\alpha-Fe$, $\varepsilon-Fe_{2.1-3.6}N$ and $\delta-Fe_2N$. Previous studies about Curie 
temperature~\cite{kano01} aren't useful because differences in Nitrogen composition lead to diverse magnetic behavior~\cite{defigueiredo95a}.
Indeed, almost any variation on Nitrogen content leads to appearance of ferromagnetic phases, with only $\delta-Fe_2N$ remaining paramagnetic.

After just few minutes of alloying it is no longer possible to distinguish between different phases by X-ray diffraction.  
The local description provided by M\"ossbauer spectroscopy cannot distinguish Fe atoms without N 
as near neighbors from $\alpha-Fe$. Even by transmission electron microscopy that separation is no longer possible~\cite{defigueiredo95a}. 
Therefore it is only possible to distinguish between ferromagnetic and paramagnetic iron in milled iron nitrides and on the magnetic point of
view, the system is composed by two phases: one paramagnetic and one ferromagnetic. Since there are no direct link between 
tribochemical activity and crystallography (tribochemical activity is a parameter to measure a given activity), this concept can be perfectly 
applied here.

Either phase is suited for the kind of analysis performed in this paper. Due to experimental convenience, the paramagnetic phase was chosen. 
The evolution of tribochemical activity to destroy the paramagnetic phase can be followed by eq.~\ref{eqTribo}: 
\begin{equation} 
\xi^{P}(t)\propto \frac{dX_{P}(t)}{dt}\frac{1}{X_{P}(t)[1-X_{P}(t)]}  \label{eqTriboFeN} \textrm{.} 
\end{equation} 
The tribochemical activity versus milling time for the Fe-N system studied is showed in fig.~\ref{figTriboNit}. The curve was calculated
using eq.~\ref{eqTriboFeN} and data in Fig.~\ref{figVolNit}.

\section{Discussions} 

\begin{figure} 
\begin{center} 
\includegraphics[angle=-90,width=\linewidth]{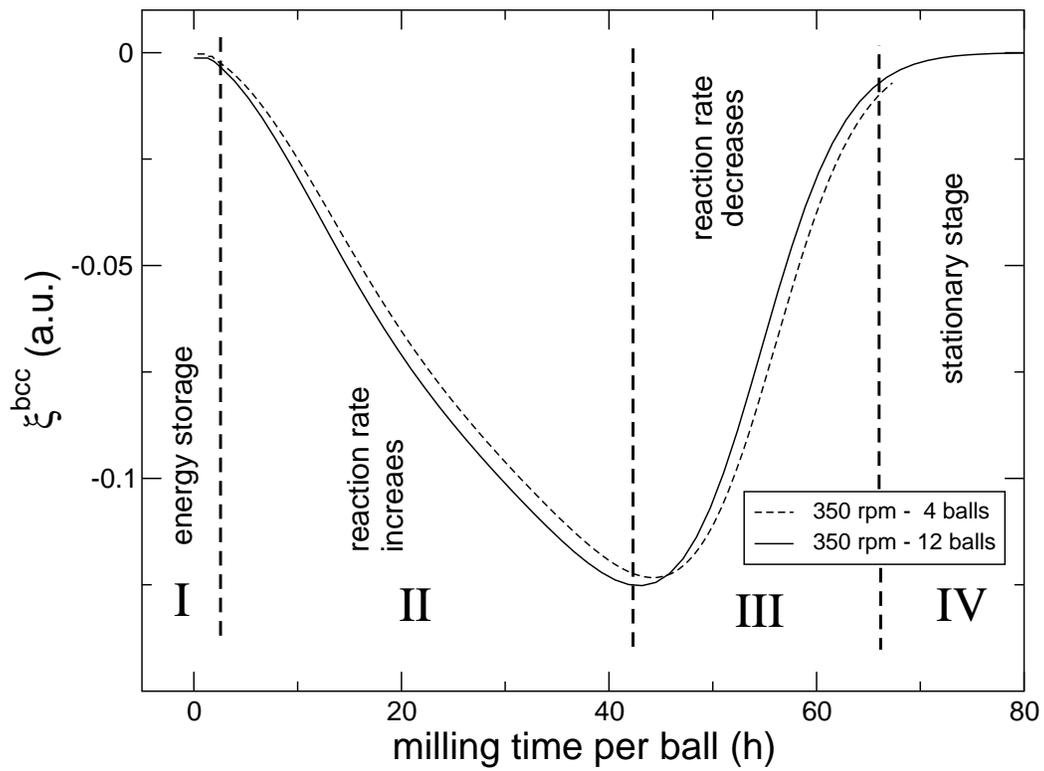}
\end{center} 
\caption{Tribochemical activity to destroy the bcc phase in the 
Fe-Cu system as a function of milling time rescaled by the number 
of balls.}\label{figTribo} 
\end{figure}

The two curves in Fig.~\ref{figTriboses} are not in the same scale because the number of balls, which influences the number of 
impacts per unit time, is different in each case. We assume the number of impacts per unit time between balls and the vial wall 
equals the number of balls in the vial. In this way, there are at least $n=3$ times as many impacts in the experiment with 12 balls. 
A correction to be made refers to the different number of impacts between balls in each experiment. It is virtually impossible to 
calculate such a number due to the nearly unpredictable dynamics of the balls in the vial. However, we empirically estimate this 
correction to be 1/5 of $n$. Therefore, there are about $n=3.6$ as many impacts in the 12 balls case. To bring both curves to scale, 
we modify the 12 balls curve by multiplying time (and dividing $\xi$) by $n$. Furthermore, we multiply the time axis (and divide 
the $\xi$ axis) by 4 to obtain tribochemical activity $vs.$ time, per ball. In this way, we get the two curves in Fig.~\ref{figTribo} 
which represent the tribochemical activity time evolution we would obtain if there were only one ball in the vial. The curves in 
Fig.~\ref{figTribo} are very similar indicating that the number of balls (impact frequency) does not play an important role on the 
final product. 

The behavior of the curves in Fig.~\ref{figTribo} reveals four distinguished stages. The first one is a static activation stage 
where the system only stores energy and no chemical reaction occurs~\cite{takacs02}. The second stage is characterized by the 
start of reaction followed by an increase of reaction speed. In fact, the reaction occurs only in the bcc/fcc interface and so this 
acceleration is expected when total surface increases (grain size decreases) like in fig.~\ref{figTamGrao}. The third stage is 
characterized by a decrease in the reaction rate. During this stage, the amount of iron dissolved in the copper matrix approaches 
the solubility limit of $\approx $60\%~\cite{huang96,uenishi92}, which explains the decrease. The fourth and last stage is 
characterized by a zero $\xi $ value revealing that stationary stage is reached. 

\begin{figure} 
\begin{center} 
\includegraphics[angle=-90,width=\linewidth]{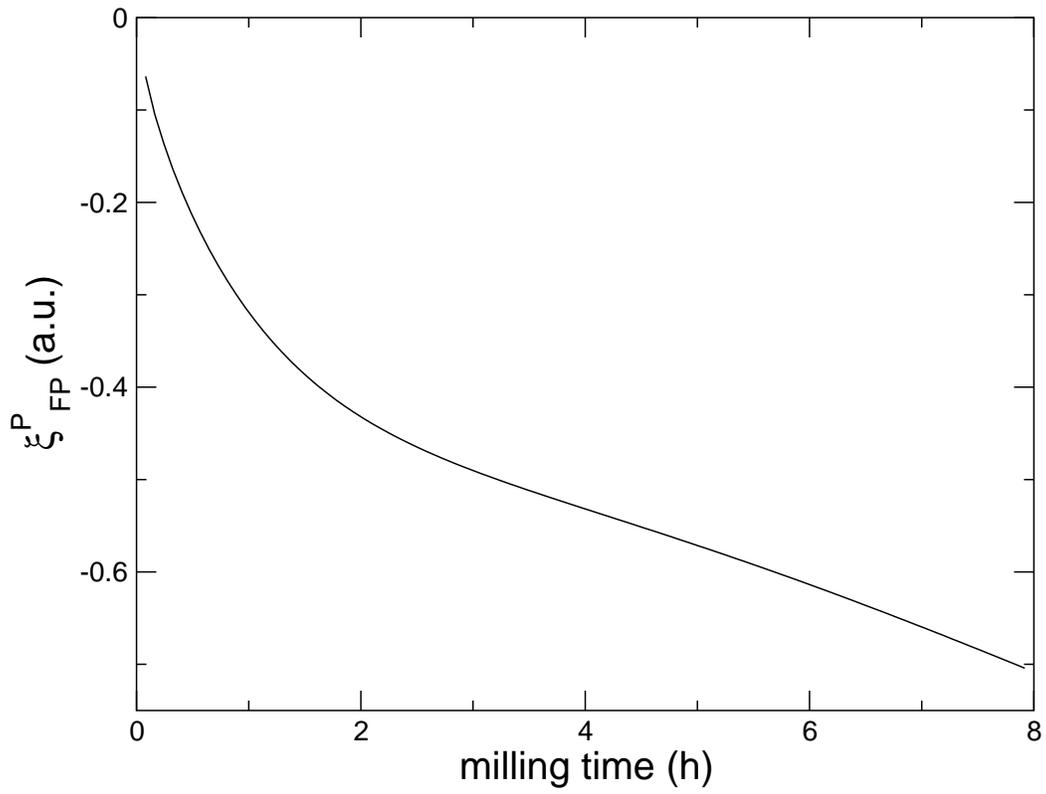}
\end{center} 
\caption{Time evolution of tribochemical activity to destroy the 
paramagnetic phase in the Fe-N system. The curve was calculated
from data in Fig.~\ref{figVolNit} using eq.~\ref{eqTriboFeN}.} \label{figTriboNit} 
\end{figure}

In contrast to what one can observe in the Fe-Cu formation process, the tribochemical activity for the Fe-N system shown in 
Fig.~\ref{figTriboNit} reveals large reactivity from the beginning of the process ($\frac{d\xi}{dt}$ is large when $t=0$). Moreover, 
the reactivity seems to be proportional to the concentration gradient between the two phases ($Fe$ and $Fe_{2}N$) and the reaction 
rate increases due to an ever growing interface between $Fe_{2}N$ and $Fe$ (grain size reduction). In fact, the Fe-Cu is an endothermic 
system and energy storage is necessary in order to trigger the reaction. This does not occur in the exothermic $2Fe+Fe_{2}N$ reaction. 
It can also be observed that the absolute value of $\xi _{FP} ^{P}$ is a monotonic time function showing that saturation has not been 
achieved, and that there was no reduction of reagents reactivity capacity. The reaction stops because of the reactants extinction. 
This is another contrast to the Fe-Cu system in which a stationary stage is reached. 

The presence of a stationary stage with zero $\xi$ in the end of the Fe-Cu processing suggests us that there might be some
unreacted amount of material left over. This fact calls for adjustment in eq.~\ref{eqmatteazzi}. Since the Fe-Cu reactions reached 
saturation, it is reasonable to believe that there should be a constant term in eq.~\ref{eqmatteazzi} to account 
for the phase volume not transformed. Thus, the modified equation takes, the form:
\begin{equation} 
V(t)=ke^{-bt^c} + R\textrm{,} \label{eqmatteazzimod} 
\end{equation} 
and plugging this expression into eq.~\ref{eqTribo}, we obtain:
\begin{equation} 
\xi_{endo}^{\varepsilon }(t)\propto - \frac{cbt^{c-1}}{1-ke^{-bt^c}-R[2-ke^{bt^c}(1-R)]} 
\textrm{.}  \label{eqKsiCoefEndo}
\end{equation}
On the other hand, the Fe-N system never reached saturation indicating that there shouldn't be any
unreacted material left over and therefore $R$ should be zero for exothermic reactions reducing eq.~\ref{eqmatteazzimod} to
the original one (eq.~\ref{eqmatteazzi}) and eq.~\ref{eqKsiCoefEndo} to:
\begin{equation} 
\xi_{exo}^{\varepsilon }(t)\propto - \frac{cbt^{c-1}}{1-ke^{-bt^c}} 
\textrm{.}  \label{eqKsiCoefExo}
\end{equation} 

It is useful to analyze the limiting behavior of eqs.~\ref{eqKsiCoefEndo} and~\ref{eqKsiCoefExo}. For large $t$, eq.~\ref{eqKsiCoefEndo} 
becomes:
\begin{equation}
\xi_{endo} \propto -\frac{t^{c-1}}{const. + (R-R^2)e^{t^c}} \textrm{.}
\end{equation}
For endothermic processes, $R > 0$ and $|\xi| \sim 1/e^{t^c} \rightarrow 0$, as expected. In eq.~\ref{eqKsiCoefExo}, the numerator
dominates and $|\xi| \sim t^{c-1} \rightarrow \infty$, which is consistent
with exothermic reactions. For small values of $t$, the denominator in eq.~\ref{eqKsiCoefEndo} and~\ref{eqKsiCoefExo} 
is constant and then the numerator is dominant. The time derivative of $\xi$ when $t$ is small goes
like $-t^{c-2}$. Therefore, for $c>2$, $d\xi/dt$ is zero in the beginning of the reaction, which is congruent
with an energy storage stage which should be present in endothermic processes. On the other hand, for $c<2$,
$|d\xi/dt|$ is maximum when $t \rightarrow 0$ which is compatible with exothermic reactions.

To test these limits, fittings of the tribochemical activity time evolution curves in Figs.~\ref{figTribo} and~\ref{figTriboNit} to 
eqs.~\ref{eqKsiCoefEndo} and~\ref{eqKsiCoefExo}, respectively, were performed. It is important to point out that eqs.~\ref{eqKsiCoefEndo} 
and~\ref{eqKsiCoefExo} are result of an empirical assumption and there is no argument to sustain that the fittings made here are, by any means,
the best ones. One could obtain fittings as good as these ones using many different expressions. However, although the aim of the analysis made
here is not to establish the absolute validity of the law in eqs.~\ref{eqmatteazzi} and~\ref{eqmatteazzimod}, it provides further insights on
the usefulness of the tribochemical activity approach.

Table~\ref{tabFit} shows the values of coefficients $k$, $b$, $c$, and $R$ obtained from numerical fitting of tribochemical activity time evolution 
curves to eqs.~\ref{eqKsiCoefEndo} and~\ref{eqKsiCoefExo} where the coefficient $R$ applies to the former but not to the latter. Comparing 
the values obtained from the two Fe-Cu 
experiments (with 4 and 12 balls) we see that the values of coefficient $b$ are quite different, which is in accordance to the proposed 
relation of them to ``work conditions''. The values of $c$ from both experiments are very similar. This was expected since 
$c$ should be connected to impact energies and both experiments were performed with rotational 
frequency of 350 rpm. As a consequence, approximately the same amount of energy is transferred to the powder 
in both experiments and the local kinetics of both reactions are the same, despite the different number of balls. 
This analysis strengthens the suggestion that the shock frequency is only a milling condition and does not 
play a role in the energy transfer process, i.e., it is not a fundamental parameter of the process. The 
values of $k$ and $R$ are virtually the same since one process differs from the other by only the time scale.
Moreover, according to tab.~\ref{tabFit}, $c$ is less than 2 for the exothermic Fe-N system in contrast to the
endothermic Fe-Cu system where $c$ is greater than 2. These numbers are in accordance to the previous discussion. 
In this way, not only is $c$ connected to impact energies as suggested in~\cite{matteazzi93}, but also reflects the 
endothermic or exothermic nature of the reaction.

\section{Conclusions} 

The formulation of the tribochemical activity concept provides a clear way to understand the 
evolution of phase transformations during MA. This is an initial attempt to describe the growth laws from 
global kinetics. Using this description it is possible to follow grain sizes, gradient of concentration and 
milling condition without chemical conditions. 

The time evolution of tribochemical activity allows the identification of four distinct stages in the milling 
of Fe-Cu system: energy storage, reaction initiation and rate increase, reaction rate decrease, and stationary.  
These stages are in perfect agreement to the experimental observations. The energy storage stage is not identified in 
the $2Fe+Fe_{2}N \longrightarrow Fe_{4}N$ reaction because of the exothermic nature of this system. Moreover, 
the monotonous character of the $\xi (t)$ function shows that saturation is not achieved. 

The striking difference between $\xi$ behavior for Fe-N and Fe-Cu reflects the respective exothermic and endothermic nature
of the reactions. Achievement of a saturation stage in Fe-Cu processes suggests the existence of some amount of unreacted 
material left over. This fact calls for modification on the global kinetics law proposed in~\cite{matteazzi93}. The small 
adjustment introduced extends the validity of the law to endothermic and exothermic processes and imposes a threshold to 
the characteristic exponent.

Further analysis of the two Fe-Cu tribochemical activity curves suggests that impact frequency does not play an important 
role in the final product process. This is stressed by the values obtained from the numerical fitting of the phase volume 
and $\xi$ time evolutions. 

At this point, it is really hard, and certainly premature, to assert that the tribochemical 
activity approach is definitive, but it definitely makes it clearer that the right path to fully understand the MA processes
should bend more and more toward its description in terms of growth laws. Such description provides some clues about which 
parameters actually drive MA mechanisms, and also provides a means to separate out surface, mechanical impacts, and thermodynamics 
terms. The rather different tribochemical activity behavior for the Fe-Cu and Fe-N systems might explain the broad diversity and 
incompatibility among various MA models present in the literature.

\newpage

\begin{table}
\begin{center}
\caption{\label{tabGrao}  Grain size for some Fe-Cu milled samples with 12 balls and 
350 rpm. This results were obtained refining the x-rays patterns 
using Rietveld method. \newline} 
\begin{tabular}{ccc}
\hline
& \multicolumn{2}{c} {\textbf{grain size (nm)}} \\ 
\cline{2-3}
\textbf{time(h)} & \textbf{bcc phase} & \textbf{fcc phase}\\
 \hline%
1 & 74 & 44 \\%
2 & 41 & 38 \\ %
4 & 33 & 24 \\ %
8 & 15 & 14 \\ %
12 & 14 & 14 \\ %
16 & - & 13 \\ 
\hline
\end{tabular}
\end{center}
\end{table}

\begin{table}
\begin{center}
\caption{\label{tabFit} Coefficients obtained by fitting Figs.~\ref{figTribo} and~\ref{figTriboNit} to 
eqs.~\ref{eqKsiCoefEndo} and~\ref{eqKsiCoefExo}, respectively. \newline}
\begin{tabular}{ccccc}
\hline
& \textbf{k} & \textbf{b} & \textbf{c} & \textbf{R} \\  
\hline          
Fe-Cu  (4 b) &  0.479 &  0.003 & 2.729 & 0.075 \\ 
Fe-Cu (12 b) &  0.485 &  0.121 & 2.658 & 0.069 \\ 
Fe-N         &  0.497 &  0.106 & 1.661 & -  \\  
\hline
\end{tabular} 
\end{center}
\end{table}

\newpage

\newpage

\newpage

\newpage

\newpage

\newpage

\newpage

\newpage

\end{document}